\def\lap{{\scriptstyle \sim}\kern-.6em\raise.8ex\hbox{$\scriptstyle < $} }
\documentclass[preprint,12pt]{elsarticle}

\usepackage{amssymb}
\usepackage{amsmath}

\journal{Annals of Physics}

\begin{document}

\begin{frontmatter}



\title{Does Dirac equation for a generalized Coulomb-like potential in D+1 dimensional flat spacetime admit any solution for $D\geq 4$?}


\author[1,2]{F. Caruso}
\ead{francisco.caruso@gmail.com}

\author[2]{J. Martins} \author[2]{L.D. Perlingeiro} \author[2]{V. Oguri}
%\author[2]{V. Oguri}
\address[1]{Centro Brasileiro de Pesquisas F\'{\i}sicas - Rua Dr. Xavier Sigaud, 150, 22290-180, Urca, Rio de Janeiro, RJ, Brazil}
\address[2]{Instituto de F\'{\i}sica Armando Dias Tavares, Universidade do Estado do Rio de Janeiro - Rua S\~ao Francisco Xavier, 524, 20550-900, Maracan\~a, Rio de Janeiro, RJ, Brazil}
\begin{abstract}
The relativistic hydrogen atom in an Euclidean space-time of arbitrary number of space dimensions ($D$) plus one time dimension is revisited. In particular, numerical solutions of the radial Dirac equation for a generalized Coulombian potential proportional to $1/r^{(D-2)}$ are investigated. It is argued that one could not find any physical solution for $D\geq 4$.

\end{abstract}

\begin{keyword}
space \sep dimensionality \sep Dirac equation \sep hydrogen atom \sep Coulombian potential.

\end{keyword}

\end{frontmatter}

% \linenumbers

\section{Introduction}
\label{int}
There are still many open questions concerning the problem of understanding how Physics depends on spacetime dimensionality. One of them is wether or not a stable hydrogen atom could exist in spaces having a number of dimensions greater than four, particulary, in the quantum-relativistic framework. A contemporary and comprehensive survey of dimensionality can be found in~\cite{Petkov}.

The idea that a particular physical law should depend on space dimensionality can be traced back to a philosophical speculation due to Kant~\cite{Kant}, namely, the suggestion that the Newtonian gravitational force depends on the three dimensionality of space. This insight was indeed proliferous and had inspired, for example, Ehrenfest~\cite{Ehrenfest17, Ehrenfest20}, who was the first to give a mathematical ground to Kant's hint. Indeed, he discussed the mechanical stability of the formal solutions of the planetary motion in $D$ dimensions, assuming that the gravitational potential is still described by a Poisson equation written in an Euclidean $D$-dimensional space $R^D$. Also, the implications of a generalized higher dimensional Coulombian potential on Bohr's atomic model was for the first time examined.

Those papers suggest that one can learn about space dimensionality from a class of generalized physical equations~\cite{Caruso}. In particular, this possibility was reconsidered in Ref.~\cite{Tangherlini}, where the problem of Schr\"odinger's hydrogen atom in $D$ dimensions was formally treated.
%The main point of Tangherlini's as well as others results which came from additional investigations concerning the stability of hydrogen atom in higher dimensions will be
%discussed in the next Section.
%Although the basic assumption that one can \textit{prove} the three-fold nature of space from the generalized Schr\"odinger's equation
%was shown to have epistemological limitations~\cite{Caruso}, the general ideas that it is possible to understand how the structure of a particular physical equation depends on dimensionality, or still how dimensionality itself can depend on the spatial scale or even how a physical phenomenon could vary by changing the number of dimensions or other topological feature of space are still of considerable interest.

%In a nutshell, the study of such a relationship between the structure of physical laws and space dimensionality is particularly attractive in the framework of the general theoretical scenario where extra space-time dimensions should play an important role on several attempts to unify the fundamental forces. Also the expectation from various scenarios of quantum gravity that the space-time dimension seems to rely on the size of the probed region, being somewhat smaller than four at small scales and monotonically raises with increasing the size of region \cite{Maziashvili} can be argued to justify the investigation on how physical laws depend on or are entangled in space-time dimensionality.

Another kind of investigation is inspired by the question of how physical laws depend on spacetime dimensionality.
As a few examples, one can quote the Casimir Effect~\cite{Bender, Rafael, Neto} and how recent data on cosmic microwave background can be used to settle an upper limit for fractal space dimensions~\cite{Oguri}.

Turning back to the problem of hydrogen atom defined in multiple dimension flat spacetimes, one realizes that, in spite of a significant literature briefly reviewed in the next Section, it is not yet clear what happens to the atomic spectrum and stability so far a higher dimensional is considered. This is particulary true in the case of the relativistic hydrogen atom. Therefore, this paper is aimed to give an answer to the following question: Does Dirac equation for a generalized Coulomb-like potential in D+1 dimensional flat spacetime admit any solution for $D\geq 4$? Somehow, its our guess that in addition to the answer it could also be possible to learn more about physics in three dimensions by investigating its generalization to spaces of higher dimensions.

\section{The two Ansatze}
\label{ansatz}

All the discussions about space dimensionality concerning any form of physical potential has to face the epistemological problem that we are not able to probe or even infer its mathematical formula from experience~\cite{Caruso}. In the particular case of the Coulombian potential, two Ansatze can be assumed. What will be called throughout the paper as Ansatz 1 admits that the $1/r$ behaviour of the three-dimensional Coulombian potential is still the same, no matter the number of space dimensionality is considered~\cite{Nieto}-\cite{Shaqqor}. Ansatz 2, inspired in an early proposal from Ehrenfest, supposes that the potential depends on the dimensionality $D$ as $1/r^{D-2}$~\cite{Supplee}-\cite{Jordan}. Ansatz 1 has the advantage to give rise to some analytical solvable problems. In addition, it is well established that a system defined through the $1/r$ potential is stable, irrespectively of the number of spatial dimensions, for both non-relativistic and relativistic cases~\cite{Dong-book}. However, Ansatz 2, in spite of an intrinsic mathematical difficulty, has the advantage of ensuring, at least at the classical level, the electric charge ($e$) conservation~\cite{Supplee}. Indeed, in $D$ dimensions, from the Poisson equation for the electric field, $\vec E = - \vec \nabla \varphi$, it follows the integral form of Gauss law
$$\int (\vec \nabla \cdot \vec E) \mbox{d}V = \int (-\nabla^2 \varphi) \mbox{d}V\ \sim e \quad \Rightarrow \quad \varphi \sim \frac{e}{r^{D-2}}  $$

Therefore, in this paper we will investigate if there are any solutions for the $D$-dimensional Dirac equation assuming Ansatz 2, and compare our result with those obtained from Ansatz 1~\cite{Dong}.

\section{Dirac equation in a D+1 dimensional flat spacetime}
\label{former_prediction}

For convenience, in this Section, we shall follow and summarize the notation of Ref.~\cite{Dong}.
The Dirac equation for a particle with mass $M$ and electric charge $e$, in natural units ($\hbar = c = 1$), can be written as
\begin{equation}\label{Dirac-eq-3}
i\sum_{\mu=0}^D  \gamma^\mu (\partial_\mu + i e A_\mu) \psi (\vec x, t) = M \psi (\vec x, t)
\end{equation}
The $(D+1)$ $\gamma_\mu$ matrices satisfy the usual Clifford algebra. In this paper, only the particular case where the zero component of the electromagnetic vector $A_\mu$  is non-vanishing and spherically symmetric, $A_0 = V_D(r)$, will be discussed. As stressed in Section~\ref{ansatz}, there is no consensus regarding the choice of $V_D(r)$ in $D$-dimensional space: often, the Coulombian potential established for $D=~3$, proportional to $1/r$, is assumed to be valid for an arbitrary $D$ as in Ref.~\cite{Dong}. Instead, as already said, we will use throughout this paper the formula corresponding to Ansatz~2, \textit{i.e.},
\begin{equation}
\label{generalized_potential}
V_{(D)}(r) = \frac{2 \Gamma(D/2)}{\pi^{(D-2)/2}}\, \frac{e_{(D)}}{(D-2)r^{D-2}}
\end{equation}
Therefore, the potential energy $U_D$ for an electron, with charge $-e_{(D)}$, that contributes to the Dirac equation is
\begin{equation}
\label{generalized_potential}
U_{(D)}(r) = - \frac{2 \Gamma(D/2)}{\pi^{(D-2)/2}}\, \frac{e_{(D)}^2}{(D-2)r^{D-2}} \equiv - \frac{\xi}{r^{D-2}}
\end{equation}
Considering the matricial structure of Dirac equation, the wave function can be written as a $D+1$ espinor, like~\cite{Dong}
\begin{equation}\label{espinor-2}
\psi \simeq\begin{bmatrix}
  r^{-1} F(r) \\
  r^{-1} G(r) \\
\end{bmatrix}
\end{equation}
and, according to the representation theory of the $SO(D)$ group, it is straightforward to get the two coupled radial equations involving the functions $F(r)$ and $G(r)$:

\begin{eqnarray}\label{FG}
 \nonumber \frac{\mbox{d}}{\mbox{d}r} G (r) + \frac{K}{r} G (r) &=& \left[ E - U_{(D)}(r) - M\right] F (r)\\
  \nonumber \ \\
  -\frac{\mbox{d}}{\mbox{d}r} F (r) + \frac{K}{r} F (r) &=& \left[ E - U_{(D)}(r) + M\right] G (r)
\end{eqnarray}
 where $K = \pm (2\ell + D -1)/2$.

\section{Solution of the radial equations for a generalized Coulombian potential}\label{our_result}

Defining the adimensional variable $\rho = 2 r \sqrt{M^2 - E^2}$, with $|E| \lap M$, the coupled equations (\ref{FG}) can be written as
\begin{equation} \label{dong3_generalizada}
\left\{
\begin{array}{l}
\displaystyle
  \frac{\mbox{d}}{\mbox{d}\rho} G (\rho) + \frac{K}{\rho} G (\rho) = \left[ -\frac{1}{2} \sqrt{\frac{M-E}{M+E}} + \frac{\xi (2\sqrt{M^2 - E^2})^{D-3}}{\rho^{D-2}}\right] F (\rho)\\
  \ \\
\displaystyle
  \frac{\mbox{d}}{\mbox{d}\rho} F (\rho) - \frac{K}{\rho} F (\rho) = \left[ -\frac{1}{2} \sqrt{\frac{M+E}{M-E}} - \frac{\xi  (2\sqrt{M^2 - E^2})^{D-3}}{\rho^{D-2}}\right] G (\rho)
\end{array}
\right.
\end{equation}

Introducing the wave-functions $\phi_\pm (\rho)$, such as
\begin{equation}\label{def-phi}
G (\rho) = \sqrt{M-E}\, \left[\phi_{+} (\rho) + \phi_{-} (\rho) \right]; \ \ F (\rho) = \sqrt{M+E}\, \left[\phi_{+} (\rho) - \phi_{-} (\rho) \right]
\end{equation}
we get, after some algebraic manipulations,
\begin{equation}
\left\{
\begin{array}{l}
\label{sistema1}
\displaystyle
  \phi_+^\prime(\rho) - \left(\frac{\tau}{\rho^{D-2}} - \frac{1}{2}\right) \phi_+ (\rho) = -  \left(\frac{K}{\rho} +  \frac{\tau^\prime}{\rho^{D-2}}\right) \phi_- (\rho)\\
  \ \\
\displaystyle
  \phi_-^\prime(\rho)  +  \left( \frac{\tau}{\rho^{D-2}} - \frac{1}{2}\right) \phi_- (\rho) = -  \left( \frac{K}{\rho} -  \frac{\tau^\prime}{\rho^{D-2}}\right) \phi_+ (\rho)
\end{array}
\right.
\end{equation}
where we have introduced the quantities
\begin{equation} \label{tau-l}
 \tau = \frac{2^{D-3}\xi E}{(\sqrt{M^2 - E^2})^{4-D}};\quad \quad \tau^\prime = \frac{2^{D-3}\xi M}{(\sqrt{M^2 - E^2})^{4-D}}
\end{equation}

%\begin{eqnarray}\label{phi_mais}
%\nonumber \displaystyle &&\left(\frac{K}{\rho} +  \frac{\tau^\prime}{\rho^{D-2}}\right) \phi^{\prime\prime}_+ + \left(\frac{K}{\rho^2} +  %(D-2)\frac{\tau^\prime}{\rho^{D-1}}\right) \phi_+^\prime + \\
%\nonumber \\
%\nonumber &+&\displaystyle \left\{ \tau^\prime (\tau^{\prime 2} - \tau^2) \frac{1}{\rho^{3D-6}} + K(\tau^{\prime 2} - \tau^2) \frac{1}{\rho^{2D-3}} + %\frac{\tau\tau^\prime}{\rho^{2D-4}} + K[(D-3)\tau - K\tau^\prime] \frac{1}{\rho^{D}} + \right.\\
%\nonumber \\
% &&\displaystyle \left.+ \left[ \frac{(D-2)}{2}\tau^\prime + K\tau\right] \frac{1}{\rho^{D-1}} - \frac{\tau^\prime}{4}\frac{1}{\rho^{D-2}} - \frac{K^3}{\rho^3} + %\frac{K}{2} \frac{1}{\rho^2} - \frac{K}{4}\frac{1}{\rho}  \right\} \phi_+ = 0
%\end{eqnarray}

It is straightforward to show that the functions $\phi_\pm$ satisfy the equation
\begin{eqnarray}\label{eq-complete-phi}
\nonumber  && \left\{ \frac{\mbox{d}^2}{\mbox{d} \rho^2} + \frac{1}{\rho}\, \frac{\mbox{d}}{\mbox{d} \rho} \pm \frac{(D-3)}{\rho}\, \frac{\tau^\prime}{K\rho^{D-3} \pm \tau^\prime} \left[ \frac{\mbox{d}}{\mbox{d} \rho} \pm \frac{1}{2} + \frac{K}{\rho} \frac{E}{M} \right] + \right. \\
\nonumber  \ \\
  && + \left[- \left.\frac{1}{4} + \frac{\tau}{\rho^{D-2}} \pm \frac{1}{2\rho} \, - \frac{1}{\rho^2}\,
  \left(K^2 - \frac{(\tau^{\prime 2} - \tau^2)}{\rho^{2(D-3)}} \right) \right]\right\} \phi_\pm (\rho)= 0
\end{eqnarray}

These are the two equations to be solved without any approximation. Since there is no analytical solutions, they will be solved by a numerical method defined in Section~\ref{results}.
%In the limit $E\simeq M$, one can write
%\begin{eqnarray}\label{eq-app-phi}
%\nonumber  && \left\{ \frac{\mbox{d}^2}{\mbox{d} \rho^2} + \frac{1}{\rho}\, \frac{\mbox{d}}{\mbox{d} \rho} \pm \frac{(D-3)}{\rho}\, \frac{\xi}{[\sqrt{2}\, K(\epsilon/M)^{1/2} \rho^{D-3} \pm \xi]}\, \left[ \frac{\mbox{d}}{\mbox{d} \rho} \pm \frac{1}{2} + \frac{K}{\rho} \right] + \right. \\
%\nonumber  \ \\
%  && + \left[- \left.\frac{1}{4} + \frac{\tau}{\rho^{D-2}} \pm \frac{1}{2\rho} \, - \frac{1}{\rho^2}\,
%  \left(K^2 - \frac{(\tau^{\prime 2} - \tau^2)}{\rho^{2(D-3)}} \right) \right]\right\} \phi_\pm (\rho)= 0
%\end{eqnarray}

Before going on, let us remark that, in the particular case $D=3$, Eq.~(\ref{eq-complete-phi}) reduces to that found in Ref.~\cite{Dong}, where the author used the $1/r$ Coulombian potential form as established in three-dimensional space, namely
\begin{equation}
\left\{ \frac{\mbox{d}^2}{\mbox{d} \rho^2} + \frac{1}{\rho}\, \frac{\mbox{d}}{\mbox{d} \rho}  + \left[- \frac{1}{4} + \frac{\tau \pm 1/2}{\rho}  - \frac{(K^2 - \xi^2)}{\rho^2} \right] \right\} \phi_\pm (\rho)= 0
\label{eq-dong}
\end{equation}
In this case, analytical solutions could be found in terms of confluent hypergeometric functions~\cite{Dong}.

Defining $\eta \equiv E/M = \tau/\tau^\prime$, $\lambda = (1 - \eta^2)$ and $A=(2M)^{D-3}\xi$ one get $\tau^{\prime 2} - \tau^2 = A^2 \lambda^{D-3}$, $\tau^\prime =A \lambda^{(D-4)/2}$. Therefore, Eq.~(\ref{eq-complete-phi}) for the $\phi_+$ solution can be written as

\begin{equation}\label{eq-complete-phi-new}
\left\{ \frac{\mbox{d}^2}{\mbox{d} \rho^2} + p(\rho)\, \frac{\mbox{d}}{\mbox{d} \rho} + q(\rho) \left[\tau^\prime - V(\rho)\right] \right\} \phi_+ (\rho) = 0
\end{equation}

\noindent where
\begin{eqnarray*}
  p(\rho) &=& \frac{1}{\rho} \left(1 + \frac{(D-3)A}{K\lambda^{(4-D)/2} \rho^{D-3}+A} \right) \\
  \ \\
  q(\rho) &=& \frac{1}{\rho^{D-2}} \left(1 + \frac{(D-3)K\lambda^{(4-D)/2}\rho^{D-4}}{K\lambda^{(4-D)/2} \rho^{D-3}+A} \right) \\
  \ \\
  V(\rho) &=& \frac{1}{\rho^{D-2}}\, \frac{s(\rho)}{q(\rho)} \\
  \ \\
  s(\rho) &=& \frac{\rho^{D-2}}{4} - \frac{\rho^{D-3}}{2} \, \left( 1 - \frac{(D-3) A}{K\lambda^{(4-D)/2} \rho^{D-3}+A} \right) + \\
  \ \\
          &+&  \left( K^2 - \frac{A^2\lambda^{D-3}}{\rho^{2(D-3)}}\right)\, \rho^{D-4}
\end{eqnarray*}

\noindent and  $\tau^\prime$ is a real parameter that depends on the energy and others constants.

A qualitative difference between the two set of equations for $\phi_\pm$ should be pointed out. In fact, in Eq.~(\ref{eq-dong}), the term that contains the energy is that proportional to $1/\rho$, while in our Eq.~(\ref{eq-complete-phi}) it appears in three different terms, namely, those with the highest power of $1/\rho$.

\section{Numerical results}\label{results}

Before solving the equation for $\phi_\pm$ in the case of the generalized potential given by Eq.~(\ref{generalized_potential}), it is convenient to define a small parameter $\epsilon$, such as
$$E = M - \epsilon.$$

At this point, Eq.~(\ref{eq-complete-phi-new}) will be solved by applying the Numerov numerical method \cite{Numerov_1, Numerov_2, Leroy, Vitor} which allows us to determine simultaneously the values of $\epsilon$ and the corresponding radial functions $\phi_\pm$. For this purpose, a specific program was developed by the authors in $C++$ language and both calculations and graphics were done by using the CERN/ROOT package.

In this paper, one search numerically for solutions of the Dirac equation just for $s$-wave state (angular momentum $\ell = 0$), allowing space dimensionality to vary between the interval $4 \leq D \leq 10$.

A general warning is necessary before presenting the numerical results. So far concerning the problem of space dimensionality, one should adopt something similar to the cosmological principle, according to which one should expect that the laws of physics, as determined in our neighborhood, are valid in all regions of the Universe and in all moments of its history, independently of the spacetime scale which is being probed. An analogous hypothesis should be assumed here as in all literature, namely, that the physical law established in $D=3$ will be valid for different values of the dimensionality $D$ and that the numerical values of the physical constants should not vary significantly. Otherwise it will be impossible to get any numerical prediction.

The present calculations depend on the numerical value of the electric charge, $e_{(D)}$, in $D$ dimensions. It will be assumed throughout this article that $e_{(D)}$ has the same value of $e$ measured in three dimensions. This should be justified from the results of ref.~\cite{Nasseri}, where it is shown that the numerical value of the generalized fine structure function for $D$ dimensions is very close to the three dimensional one, $1/137$, and also Planck constant $\hbar_{(D)}$ did not vary significantly with space dimensionality.

The energy values of the $D$ dimensional hydrogen atom, given by Eq.~(\ref{eq-complete-phi-new}), are determined by using the Numerov numerical method. The solution of the equation is constructed by successive iterations, from two subsequent points of an arbitrary interval $[a,b]$, where the solutions $\phi(\rho -\delta)$ and $\phi(\rho)$ are supposed to be known, with $\delta$ being a small quantity, the step of the iterations.  Performing a Taylor series expansion of $\phi(\rho)$ up to the fourth order, one gets the difference formula which allows the computation of the function at the next point $\rho + \delta$, that can be displayed as:

\begin{equation}
\phi(\rho + \delta) = \frac{p_{1}(\rho)\phi(\rho) - p_{0}(\rho)\phi(\rho - h)}{p_{2}(\rho)},
\end{equation}
where
$$
p_{0}(\rho) = \Bigg\{1-p (x) \frac{h}{2} + \bigg[s(x-h)+  p^\prime(x) \bigg]\frac{h^2}{12} \Bigg\}
$$
$$
p_{1}(\rho) = \displaystyle 2\Bigg\{1 - \bigg[ s(x) - \frac{p^\prime (x)}{5} \bigg]\frac{5h^2}{12}\Bigg\} \\
$$
$$
p_{2}(\rho) = \displaystyle \Bigg\{1 + p(x)\frac{h}{2} +  \bigg[ s(x+h) + p^\prime (x)  \bigg]\frac{h^2}{12}\Bigg\}
$$

For an arbitrary initial value of $\tau^{\prime}$, one may construct the solution from $a$, \textit{i.e.}, to the left of the match point, $\rho_{\mbox{\tiny match}}$, where $\tau^{\prime} = V(\rho_{\mbox{\tiny match}})$ and from $b$, or to the right of the match point. It is important to point out that an eigenvalue associated to one eigenfunction is only accepted to be a real solution when it passes by the continuity condition not only for the function but also to its first derivative.

First of all, in order to test our numerical program we tried to reproduce the energy spectrum of the hydrogen atom, for $\ell = 0$ and for $D=3, 4, 5,... 9$, calculated in Ref.~\cite{Dong}. The results are shown in Table~\ref{dong-comp}.

\renewcommand{\arraystretch}{1.3}
\begin{table}[htbp]
\vspace{-0.2cm}
 \caption{\small Ground state energies ($\epsilon$) of the hydrogen atom, as a function of space dimensionality $D$, showing
 a comparison between the analytical calculations of Ref.~\cite{Dong} and our numerical results using the same potential as in that paper.}
   \vspace*{0.2cm}
   %\caption{\small Espectro de energia dos estados fundamentais do \'{a}tomo de hidrog\^{e}nio, em fun\c{c}\~{a}o da dimens\~{a}o ($D$) do espa\c{c}o.
   %Compara\c{c}\~{a}o dos  valores calculados anal\'{\i}tica e numericamente, onde $\varepsilon  = E - m$.}
  \centering
  {\small \begin{tabular}{c|c|c|c|c}
  \hline
$D$& ~$(E/M)_{\mbox{\tiny analytical}}$ ~  & ~$(E/M)_{\mbox{\tiny numerical}}$~ &
$\epsilon_{\mbox{\tiny analytical}}$  (eV) & $\epsilon_{\mbox{\tiny numerical}}$  (eV)\\
\hline
  3  &   0.999973373968532  &  0.999973374637922   &    -13.606~ &  - 13.605~\\
  4  &   0.999988166295761  &  0.99998816642774~   &  -6.047 & -6.047\\
  5  &   0.999993343558597  &  0.999993376804255   &   -3.401 & -3.385\\
  6  &   0.999995739882606  &  0.999995722827431   &  -2,177 &  -2.187 \\
  7  &   0.999997041587069  &  0.999997035678922   &  -1.512 &  -1.515 \\
  8  &   0.999997826472985  &  0.999997825239895   &  -1.111  & -1.111 \\
  9  &   0.999998335893803  &  0.99999833256615~   &  -0.850  & -0.852 \\
  \hline
  \end{tabular}}
%\vspace{-0.2cm}
  \label{dong-comp}
\end{table}
\renewcommand{\arraystretch}{1.0}

\begin{figure}[htb!]
\centerline{\includegraphics[width=100mm]{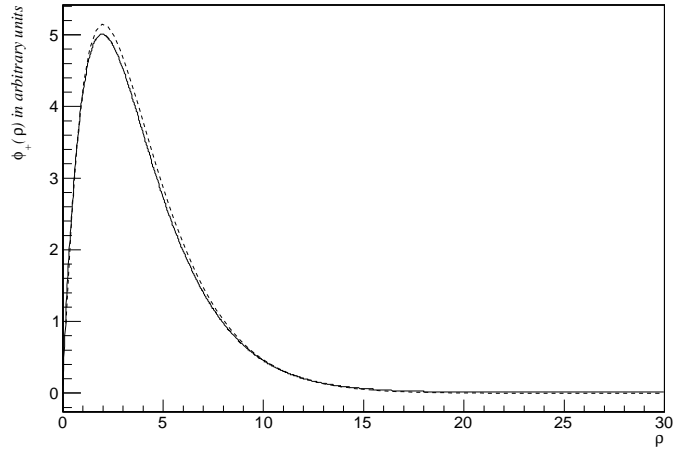}}
\caption{Comparison between our numerical result (dashed line) for the function $\phi_+(\rho)$ with the analytical one (continuous line) obtained in Ref.~\cite{Dong}, for the ground state in three dimensions.}
\label{fig.1}
\end{figure}

In addition, for $D=3$ and $\ell =0$, we compare the analytical function $\phi_+$ obtained in ref.~\cite{Dong} with our numerical result for the same function. The comparison is given in Fig.~\ref{fig.1}. Our two numerical results reproduce very well the analytical one.

\newpage

Continuing, we search for solutions of the Dirac equation, in the form of Eq.~(\ref{eq-complete-phi-new}) for higher values of $D$, always fixing $\ell =0$. For $4 \leq D \leq 10$, we were not able to find any match point that could be associated to an eigenvalue and therefore to an eigenfunction of $\phi_+$. This result corroborates those of Refs.~\cite{Supplee}-\cite{Mostepanenko} obtained in the non-relativistic case. This result was also confirmed here by using the same Numerov method~\cite{Jordan}. However, as already pointed out by Ehrenfest, in higher dimensional spaces, the atom can have positive energy levels as if the atomic electron was confined in a positive effective potential well (these will be called confined states). Indeed, we have shown that a hydrogen atom could exist as a confined state of positive energy (not as a bound state) for $D\geq 5$. These features form a consistent picture, in the non relativistic case, if we remember that there are mathematical theorems which show that considering the extra dimension to be non-compactified there should be no bound state solutions for the Schr\"odinger equation with a Coulombian potential~\cite{Bires}.

Our result (based on Ansatz 2) contrasts with the relativistic one of Ref.~\cite{Dong}, since in that paper it was claimed that, for the $1/r$ Coulombian potential (Ansatz 1), there are bound states for any space dimensionality (except $D=1$). On the one hand, it is true that there is no way to choose one or another Ansatz. However, it seems to us a little bit difficult to sustain that one should expect the hydrogen atom to have relativistic bound states which are strictly forbidden in the non-relativistic limit~\cite{Jordan,Bires}, independently of the Ansatz. By other hand, we hope that electric charge conservation still holds in $D$-dimensions.

\section{Final comments}\label{comm}

One can ask what could be learned from the negative results presented here. Admitting the hydrogen atom is described by the generalized Dirac-Coulomb equation in a $D+1$ spacetime we could not find any negative energy solution when $D \geq 4$. Although it cannot be considered a definitive proof, our numerical results suggest that the only space dimension compatible to the measured bound state energy value of $- 13.6$~eV is $D = 3$. It seems to indicate that nature should somehow prefer three-dimensionality

\section*{Acknowledgment}
One of us (J.M.) would like to thank CAPES from Brazil for financial support.

%% References


\bibliographystyle{elsarticle-num}



\end{document}